\documentclass[twoside, reqno, 12pt]{amsart}

\usepackage[left=3cm, right=3cm, top=3cm, bottom=3cm]{geometry}

\usepackage[colorlinks=true, pdfstartview=FitV, linkcolor=blue,citecolor=blue, urlcolor=blue]{hyperref}

\usepackage[usenames]{xcolor}
\definecolor{labelkey}{rgb}{0,0,1}
\definecolor{Red}{rgb}{0.7,0,0.1}
\definecolor{Green}{rgb}{0,0.7,0}
\usepackage{amsfonts, amssymb, amsmath, amsthm, mathrsfs, bbm, cjhebrew, gensymb, textcomp, mathtools, dsfont,tikz}

\usepackage[normalem]{ulem}
\usepackage{todonotes}

\usepackage{commath, setspace, subcaption, parcolumns, multirow, multicol, accents, comment, marginnote, verbatim, empheq, enumerate, stackrel, enumitem, stmaryrd }

\usepackage{lineno}

\makeatletter
\def\namedlabel#1#2{\begingroup
    #2%
    \def\@currentlabel{#2}%
    \phantomsection\label{#1}\endgroup
}
\makeatother

\newtheorem{thm}{Theorem}[section]

\theoremstyle{definition}
\newtheorem{df}[thm]{Definition}

\usepackage{hyperref}
\usepackage{hyperref}
\hypersetup{
	colorlinks = true,
	linkcolor = {red},
	citecolor = {blue}
}

\theoremstyle{remark}
\newtheorem{re}[thm]{Remark}

\newcommand{\RNum}[1]{\uppercase\expandafter{\romannumeral #1\relax}}

\numberwithin{equation}{section}

\usepackage[utf8]{inputenc}
\usepackage[english]{babel}
\usepackage{paralist}
\usepackage{amsthm}
\usepackage{marginnote}

\theoremstyle{definition}
\newtheorem{definition}{Definition}[section]
 
\theoremstyle{remark}

 \usepackage{bm,amsmath}

\newcommand{\RN}[1]{%
  \textup{\uppercase\expandafter{\romannumeral#1}}%
}

\newcommand{\Rey}{\mathcal{R}e }

\newcommand{\E}{\mathbb{E}}

\newcommand{\bff}{\mathbf{f}}
\newcommand{\bfg}{\mathbf{g}}
\newcommand{\bfu}{\mathbf{u}}

\newenvironment{Alirev}{\color{blue}}{\color{black}}
\newcommand{\AAA}{\begin{Alirev}}
\newcommand{\PPP}{\end{Alirev}}

\allowdisplaybreaks[4]

\usepackage[utf8]{inputenc}

\title[Statistical Estimates]{Statistical Estimates for 2D stochastic  Navier-Stokes Equations}
\author{Anuj Kumar\textsuperscript{1}}
\thanks{\textsuperscript{1} Email: \url{akumar241@outlook.com}}

\author{Ali Pakzad\textsuperscript{2}}
\thanks{\textsuperscript{2}Department of Mathematics, California State University Northridge, Email: \url{pakzad@csun.edu}}

\date{\today}
\subjclass[2010]{Primary 35Q30, 76F55, 35R60 ; Secondary 35Q35}
\keywords{Stochastic Navier-Stokes equations, Turbulence.}

\begin{document}
\begin{abstract}
 The statistical features of homogeneous, isotropic, two-dimensional stochastic turbulence are discussed. We derive some rigorous bounds for the mean value of the bulk energy  dissipation rate $\varepsilon $ and enstrophy dissipation rates $\chi $ for 2D flows sustained by a variety of stochastic driving forces. We show that  
$$ \varepsilon  \rightarrow 0  \hspace{0.5cm}\mbox{and} \hspace{0.5cm} \ \chi   \lesssim \mathcal{O}(1)$$ in the inviscid limit,  consistent with the dual-cascade in 2D turbulence.

\end{abstract}
\maketitle
\date{}

\section{Introduction}

A key characteristic of turbulent flows is the emergence of complex, chaotic structures across a wide range of length scales. The intricate motion makes it impractical to provide a detailed description of fluid velocity; experimental or numerical measurements of instantaneous system variables appear disordered and unpredictable. In contrast, turbulence modeling aims to predict averaged quantities rather than focusing on point-wise values. Time-averaged quantities often remain predictable, even when dynamic flow behavior is irregular over finite time intervals \cite{frisch1995}. As a result, theoretical studies of turbulence typically employ a statistical rather than point-wise description. Two relevant quantities in the  statistical  study of fully developed 2D turbulence are the energy and enstrophy dissipation rates \cite{Foias2002}. The rigorous analysis of 2D turbulence requires systematically estimating  energy and enstrophy dissipation rate as functions of externally controlled parameters, such as the Grashof or Reynolds numbers, which appropriately measure the applied stress \cite{doering1995}.

Due to the stochastic nature of turbulence and random fluctuations in the flow \cite{wilcox2006}, adding noise to the equations of motion is a common practice in both practical and theoretical applications \cite{frisch1995}. 
A central goal of this research is to understand  the effect of fluctuations, noise and randomness on key characteristics of two dimensional   turbulence- specifically, the mean value of  energy and enstrophy dissipation rates- as manifested by solutions of the Navier-Stokes equations (NSE).  We focus on the two-dimensional Navier-Stokes equation for an incompressible fluid with a random perturbation in the body forces, given by
\begin{equation} \label{SNSE}
\begin{split}
d \bfu +  ( \bfu \cdot \nabla \bfu -  \nu \Delta \bfu + \nabla p ) \, dt &= \bff \, dt  +   \bfg \, d W_t,\\
\nabla \cdot \bfu & = 0,
\end{split}
\end{equation}
within a periodic domain \( [0, \ell]^2 \) (i.e. \( \mathbb{T}_{\ell}^2 = \mathbb{R}^2/\mathbb{Z}^2 \)). In \eqref{SNSE}, stochastic processes $\bfu $ and $p$ denote the velocity field and the pressure field respectively, and $\nu$ is the kinematic viscosity. The applied force includes a deterministic part $\bff$ and a mean zero, white-in-time and colored-in-space Gaussian process $\bfg \, d W_t$ which is defined by
 \[\bfg \, d W_t= \sum\limits_{k} \bfg_{k}(x,t) dW^k(t; \xi),\] 
 where  $\{W^k(t; \xi)\}$ are a family of independent one-dimensional Brownian motions supported on a common canonical filtered probability space  $(\Omega, \mathcal{F}, (\mathcal{F}_t), \mathbb{P})$ and $\{\bfg_{k}\}$ is a family of  divergence free functions in $H^1([0,\ell]^2)$.
Equations \eqref{SNSE} have been studied as a model for fully developed turbulence in multiple situations \cite{Bedrossian2019, Bedrossian2022, FGV2016, FRT2010, CW1997, Feng24}.  In this paper,  we consider martingale
solutions (see \cite{Flandoli1995} for a detailed discussion), which are weak in both the sense of PDE theory
and stochastic analysis. The existence of martingale solutions to the stochastically forced Navier-Stokes Equations \eqref{SNSE} has been known since the 1970s with the work of Bensoussan and Temam in \cite{Temam73}.

Although 2D turbulence is much more convenient to simulate, the absence of the vortex stretching mechanism leads to an inverse cascade, which makes the phenomenology of 2D turbulence somewhat more complex than that of 3D turbulence.  Kraichnan \cite{kraichnan1967}, Leith \cite{leith1968}, and Batchelor \cite{batchelor1969} conjectured,   largely dependent upon extrapolations from what is observed in 3D and on some reasonable physical phenomenological arguments,   that there is a dual cascade in 2D turbulence: energy flows to larger scales while enstrophy moves to smaller scales. In short, in an unforced inviscid 2D flow, with the wavenumber denoted
 $k$ and the kinetic energy density $E(k)$,  the total energy  \(=  \int E(k) \, dk \)  and the enstrophy \(= \int k^2 E(k) \, dk \)  are both conserved throughout the evolution. Consequently, any transfer of energy to higher wavenumbers must be balanced by a compensating flux of energy back toward larger length scales. This characteristic of 2D turbulence is known as the inverse cascade.  On the other hand, there is a forward cascade in enstrophy, where enstrophy flows from large to small scales. This transfer occurs as larger vortices break down into smaller ones, increasing the enstrophy at smaller scales. The smaller scales then dissipate enstrophy through viscous effects.  This dual cascade mechanism is a distinctive feature of 2D turbulence, resulting in complex flow patterns where energy is concentrated at larger scales while enstrophy is dissipated at smaller scales. The interplay between these cascades contributes to the unique dynamics and statistical properties observed in 2D turbulent flows. 
\subsection{Previous work and the present results}

In the theory of turbulence, the time-averaged energy dissipation rate per unit mass,  $\varepsilon$,    and the time-averaged enstrophy  dissipation rate  per unit mass,  $ \chi$, (see Definition \ref{def_en}) represent the amount of energy and enstrophy lost to viscosity per unit mass.  These  are  two fundamental quantities  \cite{lesieur1997,  frisch1995}, determining the smallest and largest persistent length scales  in a 2D turbulence flow [see, for instance, Chapter 3 of  \cite{wilcox2006}].    Lastly, it is worth noting that estimates of the dissipation rates  have been employed to establish bounds on the dimension of the attractor for the 2D Navier–Stokes equations \cite{constantin1985, constantin1988, doering1995}.

In the context of 3D turbulence,  Doering and Constantin in \cite{DC92}   and Doering and Foias  in \cite{DF02}  proved  rigorous asymptotic upper  bounds  directly from the Navier-Stokes Equations.  Their bound is of the form  $ \varepsilon \lesssim  U^3/L$  as $\nu \rightarrow 0$.  These  works builds on \cite{Busse78, Howard72} and has developed in many important directions, e.g.  \cite{W00, AP17, L16, K16, DLPRSZ18, AP19}.  Recently the authors in \cite{FJP21} and \cite{fan20233d}  could quantify the effect of the noise by upper bounds on the first moment of the dissipation rate for  a shear turbulence flow  in the absent of external force when the fluid is driven by the noisy movement of the  boundary.   The exact dissipation rate is obtained  in \cite{ChowPakzad2022} for the 3D  stochastically forced NSE under an assumption of energy balance. In the stochastic setting, Kolmogorov's 4/3 and 4/5 laws, as well as the third-order Kolmogorov universal scaling law, have been studied in \cite{Bedrossian2019} and \cite{Dudley2024}, respectively.

The derivation of bounds for energy and enstrophy dissipation in 2D flows  has a long-standing history, as illustrated in sources such as \cite{Constantin1994, TranShepherd2002,  TranDritschel2006}.  Specifically, Foias, Jolly, Manley and Rosa \cite{Foias2002}  established a bound for the enstrophy dissipation rate in the statistically stationary states of two-dimensional turbulence, driven by a specific set of deterministic forces. Remarkably,  Alexakis and Doering in \cite{AlexakisDoering2006}  developed a systematic approach to estimate energy and enstrophy dissipation in forced flows, which can be applied to a wider variety of smoothly varying time-dependent deterministic forces.

In this letter, we establish rigorous bounds on the first moments of the dissipation and enstrophy rates in two-dimensional flows driven by various stochastic forces. We are particularly focused on the behavior of long-time averaged dissipation rates in the limit of vanishing viscosity. The
difficulty in our analysis is further compounded by the stochastic terms, along with the inviscid
conservation of both enstrophy and energy in two dimensions, which leads to two cascading quadratic invariants.  The main result of this paper can be summarized as 
\begin{thm}\label{T:Main_theorem}
    Let $\bfu$ be a martingale solution of \eqref{SNSE}, and    $\bff$ and $\bfg$ satisfy \[\sup_{t\ge 0} \| \bff\|_{H^2},\, \sup_{t \geq 0} \|\nabla \bff\|_{L^{\infty}},\,\sup_{t\ge 0} \left(\sum_{k}\| \bfg_k\|^2_{H^1} \right ) < \infty.\] Let $ \varepsilon $ and $  \chi $ denote the mean value of energy and enstrophy dissipation rates as defined in Definition \ref{def_en}.  Then, we have
    \begin{align*}
         \chi \le (\tau+1+Re^{-1})\frac{U^3}{L^3}+\tilde{G}^2,\\
         \varepsilon\le  \frac{1}{Re^{\frac{1}{2}}}\left[(\tau+1+Re^{-1})+(\tilde{G}^2 L^{3} U^{-3}\right]^{\frac{1}{2}}\frac{U^3}{ L},
    \end{align*}
where $U=$ the large-scale velocity, $L=$ the forcing length scale, $Re=$ the Reynolds number, and $\tilde{G}=$ the enstrophy rate supplied by the random force, as defined in Definition \ref{Def2}.  In particular, it follows that $\varepsilon \to 0$ and $ \ \chi   \lesssim \mathcal{O}(1)$ as ${Re}\to \infty$. 
\end{thm}
\begin{re}
Readers can refer to Page 374  Definition 3.1 of  \cite{Flandoli1995} for the detailed discussion on the martingale solution. 
\end{re}

\section{Mathematical framework and definitions}

We denote by  $(\cdot , \cdot)$, the $L^2$ inner product and by $\|\cdot \|$, the corresponding norm on  $\mathbb{T}^2$.    We recall several well-known inequalities in Banach and Hilbert spaces, which can be found in classical texts (see, e.g., \cite{Brezis2010}).

Let \( 1 \leq p \leq \infty \), and denote by \( p' \) the conjugate exponent, where \( \frac{1}{p} + \frac{1}{p'} = 1 \). Assume \( f \in L^p \) and \( g \in L^{p'} \) with \( 1 \leq p \leq \infty \). Then, the following holds
\begin{equation}\label{Holder}
 \tag{H\"older's inequality}
\|fg\|_{L^1} \leq \|f\|_{L^p} \, \|g\|_{L^{p'}}.
\end{equation}
Additionally, for any \( a, b \geq 0 \) and \( \lambda > 0 \), we have
\begin{equation}\label{Young}
 \tag{Young's inequality}
a b \leq \lambda a^p +  \left( p \lambda \right)^{-\frac{p'}{p}} \frac{1}{p'} b^{p'}.
\end{equation}

Writing $\lambda_1$ as the smallest eigenvalue of  the Stokes operator (see \cite{FMRT01}),  then for $\phi \in H^1$,  we have 
\begin{equation}\label{Poincare}
\tag{Poincar\'e inequality}
\lambda_1 \|\phi\|^2 \leq  \|\nabla \phi\|^2.
\end{equation}

In the context of stochastic calculus, Jensen's inequality (see,  for instance, \cite{Evans2013}) reads as follows: if \( X \) is a random variable and \( \phi : \mathbb{R} \to \mathbb{R} \) is a convex function, then

\begin{equation}\label{Jensen}
 \tag{Jensen's inequality}
\phi(\mathbb{E}[X]) \leq \mathbb{E}[\phi(X)]
\end{equation}
provided that \( \mathbb{E}[X] \) exists. Consequently, if \( \phi \) is concave, the inequality is reversed
\[
\phi(\mathbb{E}[X]) \geq \mathbb{E}[\phi(X)].
\]

Infinite time averaging is typically defined using the limit superior (lim sup) of a function, as
\begin{align}\label{Defn:limtavg}
\left \langle \psi(\cdot) \right \rangle = \limsup_{T \to \infty} \frac{1}{T} \int_0^T \psi(t) \, dt.
\end{align}
In some mathematical contexts, \( \left \langle \cdot \right \rangle \) represents long but finite time averages \cite{foias2005, foias2005kolmogorov, Whitehead_Doering_2012}, while in others, it refers to ensemble averages \cite{Foias2002} with respect to an invariant measure.  

\subsection*{Energy and Enstrophy dissipation rate }
In experiments, it is natural to take a long, but fixed time interval $[0,T]$ and compute the time-averages  
\begin{equation}\label{Def:timeaverageT}
\begin{split}
    \langle \mathcal{E}\rangle_T:=&\frac{1}{\ell^2}\,\frac{1}{T}\int_{0}^{T}   \nu\|\nabla u(t ,\cdot, \xi)\|_{L^2}^2  \, dt,\\
    \langle  \mathcal{X}  \rangle_T:=&\frac{1}{\ell^2}\,\frac{1}{T}\int_{0}^{T}   \nu\|\nabla \omega(t ,\cdot, \xi)\|_{L^2}^2  \, dt. \;
    \end{split}
\end{equation}
It is shown in \cite{FJMRT} that the effect of $T$ in finite-time averages of physical quantities in turbulence theory  can be controlled by parameters such as $\Rey$.   In our setting, this finite-time average in \eqref{Def:timeaverageT}  is a random variable whose mathematical expectation can be approximated by taking an average over a number of samples in the experiments. 
\begin{definition}\label{def_en}
We take the  time-averaged expected energy  and  enstrophy dissipation rate for a martingale solution $u$ of \eqref{SNSE}  to be defined by 
\begin{align}\label{varepsdef}
\varepsilon  \coloneqq \limsup\limits_{T\rightarrow\infty}\mathbb{E}[\langle \mathcal{E} \rangle_T]
=\limsup\limits_{T\rightarrow\infty}\mathbb{E}\left[ \frac{1}{\ell^2}\,\frac{1}{T}\int_{0}^{T}   \nu\|\nabla u(t ,\cdot, \xi)\|_{L^2}^2  \, dt\right]\;, 
\end{align}
\begin{align}\label{varchidef}
\chi  \coloneqq \limsup\limits_{T\rightarrow\infty}\mathbb{E}[\langle \mathcal{X}\rangle_T]
=\limsup\limits_{T\rightarrow\infty}\mathbb{E}\left[ \frac{1}{\ell^2}\,\frac{1}{T}\int_{0}^{T}   \nu\|\nabla \omega(t ,\cdot, \xi)\|_{L^2}^2  \, dt\right]\;.
\end{align}
\end{definition}

\begin{re}
By Fatou's lemma, we have
\[
\limsup_{T \to \infty} \mathbb{E}[\langle \mathcal{E} \rangle_T] \leq \mathbb{E}\left[\limsup_{T \to \infty} \langle \mathcal{E} \rangle_T \right], \quad \text{and} \quad \limsup_{T \to \infty} \mathbb{E}[\langle \mathcal{X} \rangle_T] \leq \mathbb{E}\left[\limsup_{T \to \infty} \langle \mathcal{X} \rangle_T \right].
\]
As a consequence, an upper bound on the time-averaged dissipation rates $\varepsilon$ and $\chi$, defined in \eqref{varepsdef} and \eqref{varchidef}, does not necessarily yield a bound on the expected value of their time-lim sup when the order of $\limsup$ and expectation is reversed.

\end{re}

\begin{re}
    Henceforth, we will use the convention that repeated indices indicate summation.
\end{re}

\section{Proof of Theorem \ref{T:Main_theorem}}
Before presenting the main result, it is essential to carefully define the various scales, taking into account both the domain and the underlying physics of the problem. 

\begin{df}\label{Def2}
The large-scale velocity \( U \) is defined as the square root of the expectation of the velocity

\[
U = \left \langle \mathbb{E}\left[ \frac{1}{\ell^2} \|\bfu\|^2  \right] \right \rangle ^{\frac{1}{2}}.
\]

The forcing length scale \( L \) is given by

\[
L = \min \left\{ \ell, \frac{\left \langle \|\bff\|^2 \right \rangle^{\frac{1}{4}}}{\left \langle \|\Delta \bff\|^2 \right \rangle^{\frac{1}{4}}}, \frac{\left \langle \frac{1}{\ell^2} \|\bff\|^2 \right \rangle^{\frac{1}{2}}}{\sup_{t \geq 0} \|\nabla \bff\|_{L^{\infty}}} \right\}.
\]
Using the definitions of \( U \) and \( L \), we define the Reynolds number as $\Rey = \frac{U L}{\nu}.$
Additionally, we introduce the quantities \( F \), \( G \), and \( \tilde{G} \) to characterize the magnitudes of the deterministic and stochastic components of the force. With 
$$\tilde{g}_k:=\nabla \times\bfg_{k}$$
we define
\[
F \coloneqq \left \langle \frac{1}{\ell^2} \|\bff\|^2 \right \rangle^{\frac{1}{2}}, \quad 
G \coloneqq \left \langle \frac{1}{\ell^2} \sum_{k}\|\bfg_{k}\|^2 \right \rangle^{\frac{1}{2}}, \quad 
\tilde{G} \coloneqq \left \langle \frac{1}{\ell^2} \sum_{k}\|\tilde{g}_k\|^2 \right \rangle^{\frac{1}{2}},\quad 
\]
where \( G^2 \) and \( \tilde{G}^2 \) indicates the  energy and  enstrophy rate supplied by the random force respectively. Henceforth, we will only consider the scenario where \( G , \tilde{G} < \infty \).

Next, we define the time scale \( \Omega_f^{-1} \) and the 
characteristic time $\tau$ as follows:
\[
\Omega_f = \frac{\left \langle \|\partial_t \bff\|^2 \right \rangle^{\frac{1}{2}}}{\left \langle \|\bff\|^2 \right \rangle^{\frac{1}{2}}} \quad \text{and} \quad  \tau= \frac{\Omega_f \, L}{U}.
\]
\end{df}

Before establishing the main result, we begin by demonstrating the boundedness of the associated energy quantities.
Formally, one may take the scalar product of equation \eqref{SNSE} with $\bfu$, integrate by parts in time, and apply It\^o's formula (see, e.g., \cite[Theorem 4.32]{DaPratoZabczyk2014}). This yields the following energy inequality: for any $t \in [0,T)$ and $\Delta t>0$ small, the relation holds $\mathbb{P}$-almost surely
\begin{equation}\label{EnergyEq2}
  \begin{split}
  &\|\bfu(t+\Delta t)\|^2-\|\bfu(t)\|^2 + 2  \int_t^{t+\Delta t} \nu \,  \|\nabla \bfu(s)\|^2 ds \\ &  \leq \int_t^{t+\Delta t}\sum_{k}\|\bfg_{k}\|^2 dt   +  2 \int_t^{t+\Delta t}  (\bff, \bfu(s)) ds
  + 2 \int_t^{t+\Delta t}  \big(\bfg_k(s),\, \bfu(s)\big) \, dW^k_t(s).
  \end{split}
  \end{equation}
Applying \ref{Young} together with \ref{Poincare}, we can estimate the deterministic source term as follows

\begin{align}\label{YoungaplliedEq2}
\left|\int_t^{t+\Delta t}  (\bff, \bfu(s)) ds \right| \leq \frac{1}{2\,\nu\, \lambda_1} \int_t^{t+\Delta t} \|\bff\|^2 ds + \frac{\nu}{2} \int_t^{t+\Delta t}  \|\nabla \bfu\|^2 ds. 
\end{align}

Using the above estimate and applying  \ref{Poincare} once more, we deduce from \eqref{EnergyEq2} the following

  \begin{equation}\label{EnergyEq2b}
  \begin{split}
  &\|\bfu(t+\Delta t)\|^2-\|\bfu(t)\|^2 +   \int_t^{t+\Delta t} \nu\,\lambda_1 \,  \| \bfu(s)\|^2 ds     \\&\leq   \sup_{t\ge 0} \left(\sum_{k}\|\bfg_k\|^2 \right ) \Delta t +  \frac{1}{\nu\, \lambda_1} \sup_{t\ge 0} \| \bff\|^2\,  \Delta t  
    + 2 \int_t^{t+\Delta t}  \big(\bfg_k(s),\, \bfu(s)\big) \, dW^k_t(s), 
  \end{split}
  \end{equation}
$\mathbb{P}$-a.s., for $t\in[0,T)$. Observe that by the Cauchy-Schwarz inequality and H\"older's inequality, we have
\begin{align*}\int_0^T\E\left[\sum_k (\bfg_k, \bfu)^2\right] dt &\le \int_0^T \E\left[\sum_k \|\bfg_k\|^2\|\bfu\|^2\right] dt\\&\le \left(\int_0^T\sum_k \|\bfg_k\|^2 dt\right)\sup_{0\le s\le T}\E\left[\|\bfu\|^2\right]<\infty.
\end{align*}
Hence with the standard property of It\^{o} integral \cite{Evans2013} 
$$\E \left[\int_t^{t+\Delta t}  \big(\bfg_k(s),\, \bfu(s)\big) \, dW^k_t(s) \right]=0,$$
and therefore, by taking the expectation in \eqref{EnergyEq2b} and dividing both sides by $\Delta t$, we obtain the following: for all $t \in [0, T)$
\begin{equation*}\label{EnergyEq2d}
  \begin{split}
  \frac{\E[\|\bfu(t+\Delta t)\|^2]-\E[\|\bfu(t)\|^2]}{\Delta t}+ \frac{\nu \lambda_1}{\Delta t}\int_t^{t+\Delta t}\E[ \| \bfu(s)\|^2] ds& \\\le \sup_{t\ge 0} \left(\sum_{k}\|\bfg_k\|^2 \right )+\frac{1}{\nu \lambda_1} \sup_{t\ge 0} \| \bff\|^2&.
  \end{split}
  \end{equation*}
Taking $\lim_{\Delta t \to 0}$ on both sides, we obtain for a.e. $t\in[0,T)$
\begin{align*}
    \frac{d\ \E[\|\bfu(t)\|^2]}{dt}+\nu \lambda_1\E[\|\bfu(t)\|^2]\le  \sup_{t\ge 0} \left(\sum_{k}\|\bfg_k\|^2 \right )+\frac{1}{\nu \lambda_1} \sup_{t\ge 0} \| \bff\|^2.
\end{align*}
Finally, by an application of Gr\"onwall's inequality, we obtain the following uniform-in-time bound
\begin{equation}\label{velocity:bounds}
    \sup_{t \geq 0} \mathbb{E}\left[\|\bfu\|^2\right] < \infty.
\end{equation}
Using \eqref{YoungaplliedEq2} in \eqref{EnergyEq2}, and taking the expectation, we obtain the time-averaged estimate 
\begin{equation}
    \frac{\nu}{T} \, \mathbb{E}\left[ \int_0^T \|\nabla \bfu\|^2 \, dt \right] \leq \frac{1}{T} \, \mathbb{E}\left[ \| \bfu_0\|^2 \right] + \frac{1}{T}\int_0^T\sum_{k}\|\bfg_{k}\|^2 dt+\frac{1}{\,\nu\, \lambda_1 T} \int_0^T \|\bff\|^2 ds,
\end{equation}
which ensures that the mean energy dissipation rate $\varepsilon$ is finite.    Owing to our assumption that ${\bf f(\cdot,t)}\in H^2 $ and $\bfg_{k}(\cdot,t) \in H^1$, a similar process can be carried out for the vorticity formulation \eqref{SNSE_vort} to obtain
\begin{equation}\label{vorticity:bounds}
     \sup_{t \geq 0} \mathbb{E}\left[\|\omega\|^2\right] <  \infty, \hspace{0.5cm} \text{and} \hspace{0.5cm}  \frac{1}{T} \, \mathbb{E}\left[ \int_0^T \|\nabla \omega\|^2 \, dt \right] \leq C, 
\end{equation}
which means  that the enstrophy dissipation rate  $\chi$ is also well-defined. We are now ready to prove the main result.


\subsection{Enstrophy dissipation}
The vorticity equation is derived from the curl of the velocity equation \eqref{SNSE} using standard identities. Therefore, for the scalar vorticity $\omega = \nabla \times \bfu = \partial_x u_2 - \partial_y u_1$, we have
 
\begin{equation} \label{SNSE_vort}
 d \omega +  ( \bfu \cdot \nabla \omega -  \nu \Delta \omega ) \, dt=    \phi\,  dt  +   \tilde{g}_k \, d W_t^k,
\end{equation}
where 
$\phi= (\nabla \times \bff)$ and $\tilde{g}_k=\nabla \times \bfg$.  Rewriting \eqref{SNSE_vort}, we obtain
\[d \omega =  (\phi(x,t)- \bfu \cdot \nabla \omega +  \nu \Delta \omega ) \, dt +   \tilde{g}_k \, d W_t^k.\]
Applying It\^{o}'s product rule \cite{Evans2013} and integrating in $x$, we have 
\begin{align}\label{E:Ito_vort}
    d \|\omega\|^2=2(\omega, d\omega)+\sum_{k}\|\tilde{g}_k\|^2 dt.
\end{align}
Multiplying equation \eqref{SNSE_vort} by $\omega$ and then integrating in $x$, we obtain
\[ (d\omega, \omega)+\nu \|\nabla \omega\|^2\,dt=(\phi, \omega)\,dt+(\tilde{g}_k,\omega)\,dW_t^k.\]
Using \eqref{E:Ito_vort} for the term $(d\omega, \omega)$, we obtain
\begin{align}\label{E:Ito_vort_sub}
    \frac{1}{2}d \|\omega\|^2-\frac{1}{2}\sum_k\|\tilde{g}_k\|^2 dt+ \nu \|\nabla \omega\|^2\,dt=(\phi, \omega)\,dt+(\tilde{g}_k,\omega)\,dW_t^k.
\end{align}
In the scientific literature,
$$\frac{1}{\ell^2} \|\omega\|^2= \mbox{the total enstrophy per unit mass}.$$
Integrating \eqref{E:Ito_vort_sub} in time from $0$ to $T$, dividing by $T \ell^2$, then integrating-by-parts and applying incompressibility, we obtain
 \begin{align}\label{est_ens_A}
 \begin{split}
     &\frac{1}{T \ell^2}\|\omega(T)\|^2-\frac{1}{T \ell^2}\int_0^T \sum_{k}\|\tilde{g}_k\|^2\,dt+\frac{2\nu}{T \ell^2}\int^T_0 \|\nabla \omega\|^2\,dt\\
     &=\frac{1}{T \ell^2}\|\omega(0)\|^2-\frac{1}{T \ell^2}\int_0^T (\Delta \bff, \bfu)\,dt  + \frac{1}{T \ell^2}\int_0^T  (\tilde{g}_k,\omega)\,dW_t^k.
     \end{split}
 \end{align}
By the Cauchy-Schwarz inequality, we have
\begin{align}\label{est_ens_f_cs}
    \left|\frac{1}{T \ell^2}\int_0^T (\Delta \bff, \bfu)\,dt\right|&\le \frac{1}{T \ell^2}\int_0^T \|\Delta \bff\| \|\bfu\|\,dt \notag \\&\le \left(\frac{1}{T \ell^2}\int_0^T \|\Delta \bff\|^2\,dt\right)^{\frac{1}{2}}\left(\frac{1}{T \ell^2}\int_0^T \|\bfu\|^2\,dt\right)^{\frac{1}{2}}.
\end{align}

Using the Cauchy-Schwarz inequality, H\"older's inequality, and \eqref{vorticity:bounds}, we have
\begin{align*}
    \int_0^T \mathbb{E}\left[\sum_{k}(\tilde{g}_k, \omega)^2\right]\,dt&\le \int_0^T\mathbb{E}\left[\sum_{k} \|\tilde{g}_k\|^2 \|\omega\|^2\right]\,dt\\
    &\le \left(\int_0^T \sum_{k} \|\tilde{g}_k\|^2 \,dt \right )\sup_{0 \leq t \leq T} \mathbb{E}\left[\|\omega\|^2\right]\le C_T
\end{align*}
for some positive constant $C_T$.
Hence, we use the fact that
\[
\mathbb{E} \left[ \int_0^T (\tilde{g}_k, \omega) \, dW_t^k \right] = 0,
\]
and substituting estimate \eqref{est_ens_f_cs} into \eqref{est_ens_A}, we first take the expectation with respect to $\mathbb{P}$ and then take the $\limsup$ as $T \to \infty$.  This yields

\begin{align*}
     \chi &\le \left \langle \frac{1}{2\ell^2}\sum_{k}\|\tilde{g}_k\|^2 \right \rangle+ \left \langle \frac{1}{\ell^2}\|\Delta \bff\|^2 \right \rangle ^{\frac{1}{2}}  \, \limsup_{T \to \infty}\mathbb{E}\left[ \left( \frac{1}{\ell^2\, T} \int_0^T \|\bfu\|^2 dt\right)^{\frac{1}{2}} \right]\\ 
     &  \le \left \langle \frac{1}{2\ell^2}\sum_{k}\|\tilde{g}_k\|^2 \right \rangle+ \left \langle \frac{1}{\ell^2}\|\Delta \bff\|^2 \right \rangle ^{\frac{1}{2}}  \, \left( \limsup_{T \to \infty}  \mathbb{E} \left[ \frac{1}{\ell^2\, T} \int_0^T \|\bfu\|^2  dt \right] \right)^{\frac{1}{2}},  
    \end{align*}
where the last step follows by \ref{Jensen} and the continuity and monotonicity  of the square root function.  In summary, considering the quantities given in Definition \ref{Def2},   one can  obtain the following estimate for the enstrophy dissipation rate
\begin{align}\label{est_ens_prelim}
      \chi \le \tilde{G}^2+\frac{F}{L^2}\, U.
\end{align}

Next, we estimate $F$. 
Taking the inner product in $L^2$ of \eqref{SNSE} with $\bff$, we get
\begin{align*}
   \|\bff\|^2dt+(\bff, \bfg_k)dW_t^k=(d\bfu,\bff)+\left( (\bfu\cdot \nabla \bfu,\bff)-\nu(\Delta \bfu, \bff) \right)dt.
\end{align*}
Integrating in time from $0$ to $T$, dividing by $\ell^2 T$, and taking the expectation with respect to $\mathbb{P}$, we obtain

\begin{align}\label{est_f_prelim}
     \frac{1}{\ell^2\,  T}  \int_0^T \|\bff\|^2\, dt   & +\frac{1}{\ell^2 T}\underbrace{\mathbb{E}\left[\int_0^T (\bff,\bfg_k)\,dW_t^k\right]}_{=0}=   \frac{1}{\ell^2 T} \mathbb{E}\left[\int_0^T (d\bfu,\bff) \right]\notag\\
     &   + \frac{1}{\ell^2\, T} \mathbb{E}\left[ \int_0^T   (\bfu\cdot \nabla \bfu,\bff)\, dt  \right]   -\nu  \frac{1}{\ell^2\, T} \mathbb{E}\left[ \int_0^T   (\Delta \bfu, \bff) \, dt  \right] \notag\\ 
     &:= \RN{1} + \RN{2} + \RN{3}. 
\end{align}

We now estimate the above three  terms  $\RN{1},  \RN{2}$ and $\RN{3}$ individually. Applying It\^{o}'s product rule and integrating in $x$, we obtain
\begin{align*}
    (d\bfu,\bff)=d(\bfu,\bff)-(d\bff,\bfu).
\end{align*}
Integrating in time from $0$ to $T$ and dividing both sides by $\ell^2 T$, we have 
\begin{align}\label{est_f_A}
  \frac{1}{\ell^2 T}\int_0^T (d\bfu,\bff)&=\frac{(\bfu(T),\bff(T)-(\bfu(0),\bff(0)))}{\ell^2 T}-\frac{1}{\ell^2 T}\int_0^T (\bfu,\partial_t \bff)\,dt
  \end{align}
Applying the Cauchy-Schwarz inequality, we have 
  \begin{align}   \label{est_f_cs} 
    \left|\frac{1}{\ell^2 T}\int_0^T (\bfu,\partial_t \bff)\,dt\right|\le \left(\frac{1}{\ell^2 T}\int_0^T \|\bfu\|^2\,dt\right)^{\frac{1}{2}}\left(\frac{1}{\ell^2 T}\int_0^T \|\partial_t\bff\|^2\,dt\right)^{\frac{1}{2}}.
\end{align}
By substituting the estimate \eqref{est_f_cs} into \eqref{est_f_A}, taking the $\limsup$ as $T \to \infty$, and applying Jensen's inequality, along with the continuity and monotonicity of the square root function and the time-averaging scale given in Definition \ref{Def2}, we obtain

\begin{align*}
    |\RN{1}| &\le   \limsup_{T \to \infty}\mathbb{E}\left[ \left( \frac{1}{\ell^2\, T} \int_0^T \|\bfu\|^2 dt\right)^{\frac{1}{2}} \right]\, \left \langle\frac{1}{\ell^2}\|\partial_t\bff\|^2 \right \rangle^{\frac{1}{2}}\notag\\
    & \le  \, \left( \limsup_{T \to \infty}  \mathbb{E} \left[ \frac{1}{\ell^2\, T} \int_0^T \|\bfu\|^2  dt \right] \right)^{\frac{1}{2}}\,     \Omega_f \, \left \langle \frac{1}{\ell^2}\|\bff\|^2\right \rangle^{\frac{1}{2}}. 
\end{align*}

Thus, from above  we have the following estimate for term  $\RN{1}$
\begin{align}\label{est_I}
    |I|\le {U\,  \Omega_f}\, {F}=\frac{U^2\,  \tau}{L}F.
\end{align}
To estimate $\RN{2}$, we observe that

\begin{align}\label{est_II}
    |\RN{2}|=\left| \frac{1}{\ell^2\, T}\mathbb{E}\left[ \int_0^T (\bfu \otimes \bfu,\nabla\bff) \, dt \right]\right|& \le\,  \|\nabla \bff\|_{L^\infty_t L^\infty_x}\left| \frac{1}{\ell^2\,  T}  \int_0^T \mathbb{E}\left[ \|\bfu\|^2 \right]  \, dt \right|\notag\\
    & \le \frac{U^2}{L}F,  \hspace{1cm} \text{as     }  T \to \infty.
\end{align}

Finally, we estimate term $\RN{3}$ by integrating by parts, applying the Cauchy–Schwarz inequality, and using Jensen's inequality, which gives

\begin{align}\label{est_III}
    |\RN{3}|&\le \left| \frac{1}{\ell^2\, T }\mathbb{E}\left[ \int_0^T ( \bfu,\Delta \bff) \, dt  \right  ]   \right|
     \le  \frac{1}{\ell^2\, T }\mathbb{E}\left[ \int_0^T \| \bfu \|\|\Delta \bff\|\, dt  \right]\notag\\
     &\le \left( \frac{1}{\ell^2\, T } \int_0^T\|\Delta \bff\|^2\, dt \right)^{\frac{1}{2}} \, \left( \frac{1}{\ell^2\, T } \int_0^T \mathbb{E}\left[ \|  \bfu\|^2 \right] \, dt \right)^{\frac{1}{2}}  \notag\\
     &\le \frac{F}{L^2}\, U,  \hspace{1cm} \text{as } T \to \infty. 
\end{align}

Taking  $\limsup$ from \eqref{est_f_prelim},  and  applying  estimates in \eqref{est_I}, \eqref{est_II} and \eqref{est_III},  we obtain
\begin{align*}
F^2\le  \frac{U^2 \tau}{L}\, F+\frac{U^2}{L}\, F+\nu\frac{U}{L^2}\, F.
\end{align*}
which is equivalent to  
\[F\le (\tau+1+\frac{\nu}{UL })\, \frac{U^2}{L}. \]
Using the above  estimate in \eqref{est_ens_prelim}, we obtain 
\begin{align}\label{est_ens_final}
      \chi \le (\tau+1+Re^{-1})\frac{U^3}{L^3}+\tilde{G}^2.
\end{align}
\subsection{Energy dissipation}
Using integration-by-parts along with the Cauchy-Schwarz inequality, we obtain

\begin{align*}
    \varepsilon^2=\frac{\nu^2}{\ell^4} \left \langle\mathbb{E}\left [ \|\omega\|^2\right]\right \rangle^2&=\frac{\nu^2}{\ell^4}\left \langle\mathbb{E}\left[ (\bfu, \nabla \times {\bf k}\omega)\right]\right \rangle^2\\
    &\le \frac{\nu^2}{\ell^4}\left \langle\mathbb{E}\left[ \|\bfu\| \|\nabla \times {\bf k} \omega\|\right]\right \rangle^2\\
    &\le \frac{\nu^2}{\ell^4}\left \langle\mathbb{E}\left[ \|\bfu\|^2\right]^{\frac{1}{2}} \mathbb{E}\left[\|\nabla  \omega\|^2\right]^{\frac{1}{2}}\right \rangle^2\\
    &\le \frac{\nu^2}{\ell^2}\left \langle\mathbb{E}\left[\frac{1}{\ell^2} \|\bfu\|^2\right]\right \rangle \left \langle\mathbb{E}\left[ \|\nabla \omega\|^2\right]\right \rangle=\nu U^2 \chi.
\end{align*}

Using the bound for $\chi$ obtained in \eqref{est_ens_final}, we obtain 
\begin{align*}
    \varepsilon
    ^2&\le \nu U^2 \left[(\tau+1+Re^{-1})\frac{U^3}{L^3}+\tilde{G}^2\right]\\
    &=\left[(\tau+1+Re^{-1})+(\tilde{G}^2 L^3 U^{-3})\right]\frac{U^6}{Re L^2}.
\end{align*}
Therefore, we have
\begin{align}\label{est_en_final}
    \varepsilon\le  \frac{1}{Re^{\frac{1}{2}}}\left[(\tau+1+Re^{-1})+(\tilde{G}^2 L^{3} U^{-3}\right]^{\frac{1}{2}}\frac{U^3}{ L},
\end{align}
thus completing the proof. 
\section{Discussion}

In this work, we analyzed the two-dimensional stochastically forced Navier-Stokes equations and derived upper bounds for the mean values of the time-averaged energy and enstrophy dissipation rates, which are consistent with the dual-cascade framework proposed by Kraichnan, Leith, and Batchelor.  Establishing a uniform-in-time bounds on the energy quantities  $\mathbb{E}[\|u\|^2]$  and  $\mathbb{E}[\|\omega\|^2]$ are  crucial component of the analysis. In the case of additive noise, where \( g = g(t) \) in \eqref{SNSE}, such a bound can be derived without imposing any additional assumptions. However, we conjecture that the results presented in this manuscript may also be extended to the setting of general multiplicative noise of the form \( g = g(t, u) \), potentially under a smallness condition on the diffusion coefficient (see \cite[Theorem 4.1]{Flandoli1995}).

One significant challenge arises in the estimation of the enstrophy dissipation rate. Specifically, the term $\tilde{G}^2$, which represents the enstrophy rate supplied by the random forcing, appears in the final estimation \eqref{est_ens_final}. Although $\tilde{G}$ plays a crucial role in understanding the dynamics, a more comprehensive estimation of this quantity in terms of the characteristic scales $U$ (velocity) and $L$ (length) would strengthen the results and provide deeper insights into the interplay between forcing and dissipation. Future work will focus on addressing this limitation, potentially leading to a more refined characterization of the enstrophy cascade in the stochastic setting.

\section*{Acknowledgment}
The authors would like to thank the referees for their valuable suggestions to improve the manuscript. A.P. was partially supported by NSF grant DMS-2532987.

\bibliographystyle{plain}
\bibliography{references.bib}

\hfill

\end{document}